\documentclass{article}
\usepackage{color}
\usepackage{amssymb,amsthm}
\usepackage{a4wide}
\usepackage[intlimits]{amsmath}
\usepackage{setspace}
\usepackage{paralist}
\usepackage{graphicx}
\usepackage[T1]{fontenc}
\usepackage{subcaption}			
\usepackage{lscape}				
\usepackage{tabularx}			
\usepackage[misc]{ifsym}		
\usepackage{booktabs}			
\usepackage{graphicx}			
\usepackage{multirow}			
\usepackage{ltablex}
\usepackage{longtable}
\usepackage[left=3.3cm, right = 3.3cm, bottom = 3.5cm, top = 3cm]{geometry}
\usepackage{authblk}
\usepackage{natbib}

\makeatletter
\def\ps@pprintTitle{%
 \let\@oddhead\@empty
 \let\@evenhead\@empty
 \def\@oddfoot{}%
 \let\@evenfoot\@oddfoot}
\makeatother 

\begin{document}

\title{The Future of Employment Revisited:\\ How Model Selection Determines Automation Forecasts}

\author[1]{Fabian Stephany\Letter}
\author[2]{Hanno Lorenz}
\affil[1]{\footnotesize Oxford Internet Institute,\\
fabian.stephany@oii.ox.ac.uk, \\
ORCID ID 0000-0002-0713-6010}
\affil[2]{\footnotesize Agenda Austria, T\"urkenstra\ss{}e 25, 1090 Vienna,\\
ORCID ID 0000-0003-4577-4117}

\maketitle
\begin{abstract}
\singlespace{

The uniqueness of human labour is at question in times of smart technologies. The 250 years-old discussion on technological unemployment reawakens. Prominently,  \citet{frey2017future} estimated that half of US employment will be automated by algorithms within the next 20 years. Other follow-up studies conclude that only a small fraction of workers will be replaced by digital technologies. The main contribution of our work is to show that the diversity of previous findings regarding the degree of job automation is, to a large extent, driven by model selection and not by controlling for personal characteristics or tasks. For our case study, we consult experts in machine learning and industry professionals on the susceptibility to digital technologies in the Austrian labour market. Our results indicate that, while clerical computer-based routine jobs are likely to change in the next decade, professional activities, such as the processing of complex information, are less prone to digital change.\footnote{The authors would like to thank Monika K\"oppel-Turyna, Harald Oberhofer and Wolfgang Nagl for their support and helpful comments.}\\


}
\end{abstract}

\doublespace

\newpage
\section{Introduction}\label{sec:introduction}
The motivation behind our work is the discussion about technological unemployment, which has accompanied technological processes throughout the last 250 years. The debate about the susceptibility of human labour to digital technologies accelerated since a prominent study by \citet{frey2017future} concluded that half of US employment will be automated within the next 20 years, which would pose a sizeable thread to societal stability \footnote{In addition and interaction with other global dynamics, such as rising income inequalities \citep{stephany2017your,stephany2019deepens} or climate change and mass migration \citep{hoffmann2019quantifying}.}. Their estimations are the basis for several follow-up studies, which infer that the share of jobs at risk is much smaller. Our work examines the reason for the stark diversity in previous findings about job automation. We propose that differences in the degree of susceptibility emerge mainly from model selection. In order to test this assumption, we conduct a case study similar to Frey and Osborne with a survey among Austrian research and industry experts
. Our model testing confirms that differences in previous findings on the automation of jobs are mainly driven by the design of the model, rather than heterogeneity among tasks within occupations. Our results indicate that, while clerical computer-based routine jobs are likely to change in the next decade, professional activities, such as the processing of complex information, are less prone to digital change.\\

Machines have both complemented and competed with human labour in the past. Inventive ideas and creative destruction, as \citet{schumpeter1942creative} puts it, have competed with powerful social and economic interest over the technological status quo. Various movements, such as the \textit{Luddites}, who destroyed new machinery in the 18$^{th}$ century textile industry, have tried to deter progress in times of rising unemployment. However, the \textit{Luddite fallacy} has found its way into the literature, as employment has not been eradicated alongside fast technological development, but instead continued to expanded. Rather than eliminating human labour as such, technological advancements have changed a number of work profiles and led to the creation of new professions. \\ 

Whenever modern society experiences technological advancement, concerns about technologically induced unemployment arise. In recent history, technological progress has often been linked to a displacement in specific professions \citep{bresnahan1999computerisation} or even entire industries \citep{charles2013manufacturing,jaimovich2012trend}. However, to date, technological progress has not caused mass unemployment. We have seen a shift in labour from the agricultural sector to manufacturing branches, and further into the service sector \citep{david2015there}. Overall employment has been steadily increasing worldwide, despite (or perhaps because of) technological progress. Hence, new technologies display two opposite effects on employment \citep{aghion1994growth}. On the one hand, technologies substitute human labour in order to decrease production costs and increase productivity. This displacement effect lowers employment. On the other hand, reduced production costs increase real income and hence demand. The latter effect fosters production and demand for labour.\\ 

According to \cite{goldin1998origins}, technological progress led to the simplification of work processes in the 19th century. A combination of machines and unskilled labour substituted skilled labour and decreased demand in terms of skills. However, as technologies improved, technological job displacement shifted away from skilled to unskilled labour. \cite{acemoglu2017robots} calculate that an increased use of robots in the US economy between 1990 and 2007 had a negative effect on the labour market. According to their calculations, an increase in the number of industrial robots by one, per 1,000 people employed, reduces the employment-to-population ratio by 0.18 to 0.34 percentage points. \\

Similar to signs of competition with rooters for physical work, \cite{mcafee2014second} emphasize that computerization has now started challenging human performance in cognitive tasks. \cite{beaudry2016great}, in an empirical analysis, find evidence that the demand for skilled labour has been declining in recent years. This is an indication that skills under pressure of substitution are altering as technological progress persists. \cite{david2013growth} show that the implementation of computer-based technologies has put pressure on wages. As routine tasks are increasingly automated, displaced workers reallocate to the lower skilled service sector with deteriorating wages. According to \cite{goos2009job}, this has resulted in the increased polarization of the labour market in a number of developed economies \citep[see also][]{dustmann2009revisiting}. Increasing demand for well-paid jobs in which non-routine cognitive tasks are performed, as well as non-routine manual work at the lower end of the income distribution, in combination with the automation of repetitive cognitive skills, is forcing employment away from the middle of the income distribution \citep[see also][]{autor2003skill,david2013task,michaels2014has}. \\ 

Recent publications, such as \cite{ford2015rise}, raise concerns that "this time it could be different" and there will be no room for creating new jobs. \cite{frey2017future} set the starting point for a series of papers that attempts to calculate the impact of digital technologies on the demand for human labour. Based on their original data, collected during a workshop involving machine learning experts, several papers about the susceptibility of jobs have been published.\\

Yet, transferring the data on susceptibility from the US labour market to European economies is challenging in many respects. Until now, there has been no piece of research that has analysed the impact of computer automation on the labour market by using newly collected data from European countries. This approach allows us to correct the shortcomings in transferring the original US data (O*NET) of \cite{frey2017future} to the International Standard Classification of Occupations (ISCO). It also adjusts for regional particularities in labour markets, for example, differences in regulation or cultural particularities. Even though technological innovations have become market-ready, customers may hesitate to substitute them for human interaction. In addition, we analyse the possibility of a non-linear relationship between education and future digitalization, since both low- and high-skilled jobs are assumed to be less affected by digital technologies than medium-skilled professions \cite{dustmann2009revisiting}.\\ 

Addressing previous limitations, we assume that the strong differences in the degree of susceptibility between \cite{frey2017future} and follow-up studies is due to model selection. As case study, our investigation examines the degree of future digitalization of job profiles in Austria. We link expert opinions with individual data from the OECD's PIAAC data, which in turn allow for heterogeneity among workers within the same occupation. Our results indicate that, models with a binary outcome, as applied by \cite{frey2017future} result in a much higher share of jobs at risk than models with a fractional dependent variable, as used by the OECD. In both settings, clerical computer-based routine jobs are likely to change in the next decade, professional activities with the processing of complex information are less prone to digital change. The following section \ref{sec:method} describes the methodology and data, followed by the \ref{sec:results} section, which summarizes the results, while the last section concludes the paper.\\

\section{Data and Methods}\label{sec:method}

\cite{frey2017future} were the first to attempt to quantify the potential of computer-based job displacement in the near future. Based on the estimates of robotic experts, the authors calculated the susceptibility to computerization of different jobs, according to the O*NET database in the US. They conclude that 47\% of the jobs in the US are at a high risk (>70 \% probability) of being replaced due to computerization. \cite{bowles2014computerisation} applies the same method and transfers the results to European economies using the differences in the sectoral structure of each country. He concludes that 54\% of jobs in Austria have a high risk of being displaced by computers. \\

\cite{arntz2016risk} emphasize that the method used by \cite{frey2017future} overstates the share of jobs susceptible to computerization. As \cite{frey2017future} do allow for heterogeneity in tasks between different jobs, they do not allow for alterations in the tasks within one occupation. According to \cite{arntz2016risk}, one profession may contain different sets of tasks, and thus the risk of computerization could vary within this profession. Using PIAAC survey data, they combine information about the composition of tasks within each job profile with information from robotic experts on the susceptibility of jobs for the US labour market. They further transfer the results to other OECD member countries, indicating that 9\% of US workers and 12\% of Austrian workers are at high risk due to computerization.\footnote{\cite{bonin2015ubertragung} use a similar approach for Germany, \cite{pajarinen2014computerization} for Finland, and \cite{nagl2017digitalisierung} for the Austrian economy. According to \cite{nagl2017digitalisierung}, 9\% of Austrian workers have a high risk of being automated.} Among OECD countries, Austria, as well as Germany, displays the highest share of the workforce at a high risk of computerization.\\ 

For the German labour market, \cite{dengler2015folgen} relate the risk of job automation to the tasks that are characteristic of each profession. They compute the share of tasks that can be classified as routine based, according to the classification by \cite{spitz2006technical}. According to their findings, 15\% of German workers are employed in jobs with a high risk of automation. Likewise, for Austria, \cite{peneder2016osterreich} find that 12\% of Austrian workers primarily perform routine-based tasks. \\

Similar to the approach by \cite{frey2017future}, we begin our analysis with expert opinions. Between $7^{th}$ December 2017 and $7^{th}$ January 2018, we consulted Austrian industry experts and machine learning professionals. The final data set contained 35 individual experts' opinions, with 14 individuals representatives of Austrian companies in the fields of construction, consulting, insurance, investment, media, real estate and retail, and 21 responses were from industry and academic experts in machine learning and AI. Experts from both groups were individually requested to participate in an online survey. In comparison, the expert workshop by \cite{frey2017future}, which was held in 2013 at Oxford University's Engineering Sciences Department, included 70 machine learning experts \citep{brandes2016opening}. Together with their team of experts, \cite{frey2017future} initially labelled 70 out of 703 US jobs. These binary labels were then used to predict risks of automation for all US professions. The resulting estimations formed the basis of the aforementioned studies in a European context. However, for the estimation of impacts of digital technologies on the Austria labour market, our expert opinions are better suited than the opinions stemming from the Oxford seminar. Machine learning experts are familiar with the scientific principles of the technologies disrupting the labour market, but they may not be fully aware of the social environments in which smart technologies could be implemented. For example, even when chatbots in the financial service sector become market-ready, from a technological point of view, some customers will still prefer interaction with a human. In addition, the gap between technological readiness and implementation varies to a sizeable extent between countries and cultural backgrounds. In order to address this aspect of the application of new technologies, we consulted Austrian experts from the field of machine learning/AI and professionals from various industry domains.\\

The participants in our survey were asked about their opinion on the 100 most common professions in Austria, as listed in Table \ref{tab:top100}. In contrast to the focus on the susceptibility to computerization \citep{frey2017future}, we asked our experts: \textit{"Do you think that the tasks, which are characteristic of this profession today, will be substituted, to a significant degree within the next 10 years, by algorithmic technologies (such as machine learning, computer vision and natural language processing) or mobile robotics?" (Yes=1/No=0)}. This question analyses the degree to which the nature of certain professions is going to change due to technological advancement. Answers to this question do not necessarily reflect the risk of occupations being fully substituted by technologies.\\

Experts were allowed to avoid answering the question in relation to as many jobs as they wished. However, in the end, only a small minority of jobs remained unlabelled. In order to extract an indicator of future digitalization that is unique to each profession, we calculated three measures: the \textit{mean} and \textit{mode} of all expert opinions, as well as an indicator of the experts' \textit{consensus} on each profession. The \textit{consensus} is equivalent to the mode, but only for those professions to which at least 75\% of all experts attributed the same label. With this definition of \textit{consensus}, 45 professions remained and received a binary label, as shown in Table \ref{tab:top100}.

\begin{center}
TABLE \ref{tab:top100} ABOUT HERE
\end{center}

In the second step, the profession labels were matched with profession groups from the Austrian and German samples of the 2015 OECD survey of the PIAAC. The PIAAC survey supplied our analysis with individual characteristics, as well as job- and firm-level indicators. In addition, the survey contains information about the frequency of specific tasks performed by interviewed individuals during their average working routine. These tasks, as listed in Table \ref{tab:tasklist}, include human interaction, IT usage, physical work, problem-solving, reading or understanding, and writing or calculating. As the individuals provided answers about the frequency by which they undertake a given task, we normalized the answers according to the value of the working hours as follows: \textit{'on a daily basis'} (value=1), \textit{'less than daily, but more than once a week'} (value=1/2), \textit{'less than once a week, but more than once a month'} (value=1/7), \textit{'less than once a month'} (value=1/30), or \textit{'never'} (value=0). This labelling is likewise applied by \cite{arntz2016risk}, since it reflects the differences in scale between days, weeks and months. 

\begin{center}
TABLES \ref{tab:tasklist} ABOUT HERE
\end{center}

Thirdly, the expert opinions about the future change of professions are related with the PIAAC data. These opinions about professions are matched via the ISCO-08 classification for each individual's job\footnote{Only the German PIAAC sample contains the respective ISCO-08 Level 4 job classifications. Hence, the fitting of the inferential models is performed only with the labelled subset of the German employees.}. The PIAAC survey is conducted in a way that it contains a representative sample of the population. However, not all observations within the survey contain answers to all questions. Thus, the specification of the model leads to a loss in observations due to non-responses. There is no reason to assume that the loss in observation systematically changes the sample. We perform a mean imputation for the non-response values, which increases the model's sample size by 55\%, but does not lead to a significant difference in results. Compared to the 2012 labour force survey, our sample displays a slight shift towards younger age groups. Furthermore, the sample shows a higher share of female employees (for details, see Table \ref{tab:summary}). Nevertheless, the impact of technological change on job profiles stays unchanged.\\ 

In order to relate the above-mentioned characteristics to the given expert opinions about the individual's job, we test three inferential models. The \textit{consensus} indicator serves as the dependent variable, while various combinations of personal-, job- and firm-level controls, as well as task frequencies, are included in the model (Table \ref{tab:summary}). The correlation analysis in Table \ref{tab:correlation} across all characteristics only indicates a sizeable association between the three test score variables. All measures are considered at the individual level with a sample of 507. The extrapolated sample contains 4,438 individuals: 2,051 from Austria and 2,387 from Germany. In a first round, we apply a logit model. This stepwise procedure is illustrated in Columns (1)-(6) in Table \ref{tab:modelcompare}. The Akaike information criterion indicates that Model (6), with all controls, yields the best model fit. In the second round of the model selection, we test a linear discriminant analysis (LDA) with a Bayesian estimation of the dependent variable \cite[Chapter 4]{james2013introduction}\footnote{The probability of belonging to class k, given characteristics X, is described by $P(Y=k|X=x) = \frac{f_{k}(x) \pi_{k}}{P(X=x)}$, \hspace{0.05cm}, while $f_{k}(x)$ describes the probability of $X=x$, given that $Y=k$, while $\pi_{k}$ is the prior probability of observing $Y=k$.}, which is similar to the approach chosen by \cite{frey2017future}. In order to compare the logit and LDA models, we apply a cross-validation method (40\% training sample). The comparison of the in-sample predictions shows that the logit model (area under the curve (AUC)\footnote{The AUC measures the area under the receiver operating characteristics (ROC) curve. The AUC is a measure of prediction accuracy, since the ROC curve plots the true positive rate against the false positive rate of a prediction model.}: 0.94) slightly outperforms the LDA model (AUC: 0.92). The estimations of the LDA model are very similar to the results of the logit model, as summarized in Table \ref{tab:comparison}. Lastly, we compare the results of the logit model with a fractional response model \citep{papke1993econometric}\footnote{$E(y|X) = \frac{e^{(\beta'X)}}{1+e^{(\beta'X)}}$, \hspace{0.05cm}, while $\beta'X = \beta_{0} + \beta_{1}x_{1} + \cdots + \beta_{k}x_{k}$}. In this model, the mean of the experts' opinions is considered as the dependent variable. Accordingly, the fractional model refers to a larger sample size. However, the results in Table \ref{tab:modelcompare}, Columns (6) and (7), show that the logit model still yields a significantly better model fit.

\begin{center}
TABLE \ref{tab:summary} AND \ref{tab:modelcompare} ABOUT HERE
\end{center}

\begin{figure}[h!]
\begin{center}
\includegraphics[width=0.9\linewidth, keepaspectratio, ]{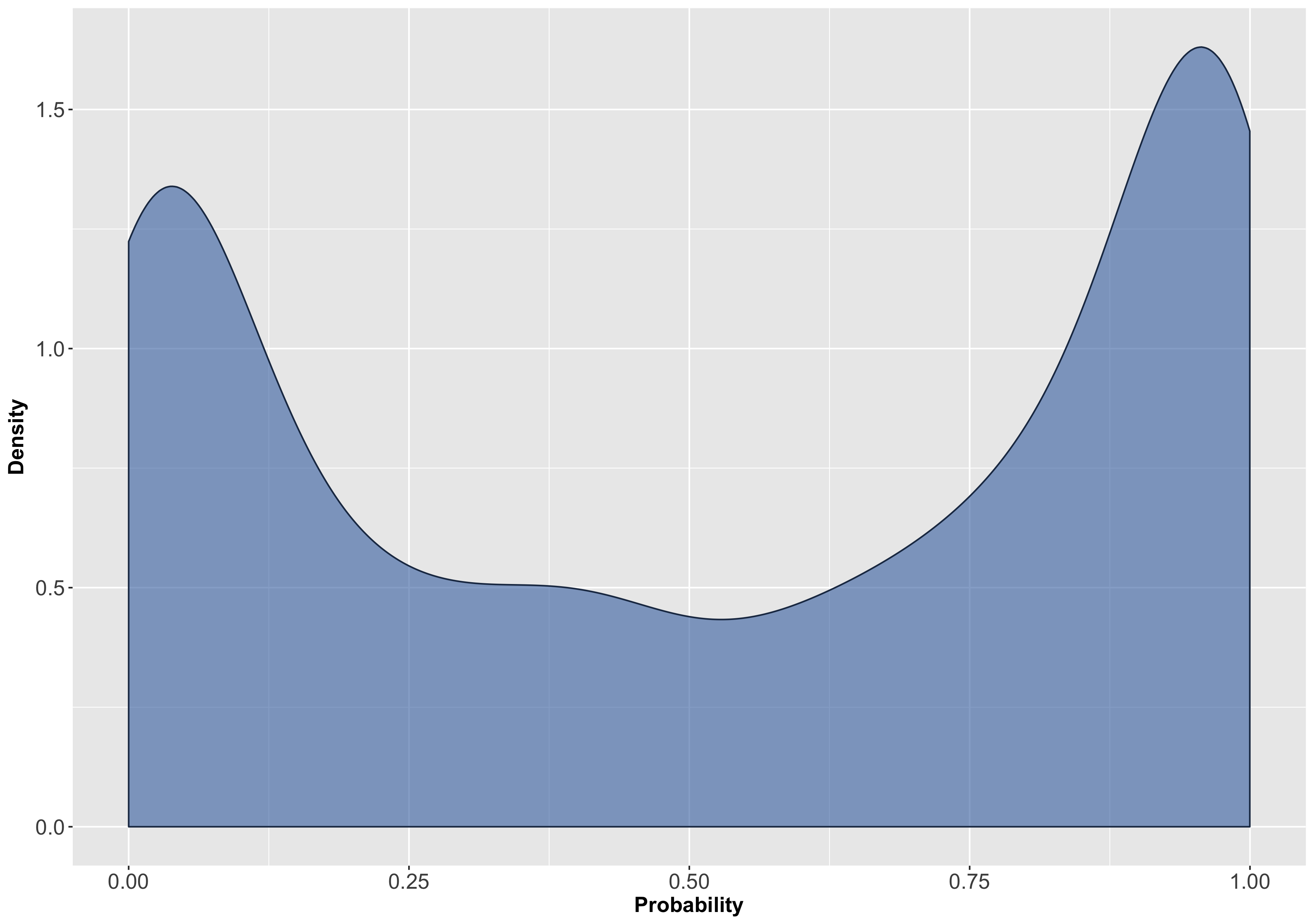}
\caption{\emph{Future digitalization}: Jobs in Austria are polarized between high and low levels of future digitalization. The distribution of individual levels of future digitalization mirrors the initial estimation of our experts.}
\label{fig:distribution}
\end{center}
\end{figure}

After identifying the appropriate model environment, the logit model (1) is used to predict the digitalization probabilities, P(y=1|X), for all individuals in the sample, based on their set of characteristics ($\beta'X$). Here, individuals with professions, which have not been judged by our experts, also obtain a probability. The average estimated probabilities of future digitalization are shown in Figure \ref{fig:distribution}, and are aggregated for ISCO-08-Level 1 (Figure \ref{fig:isco_level1}) and ISCO-08 Level 2 (Figure \ref{fig:isco_level2}) professions in Austria.

\begin{center}
$P(y = 1 | X) = \frac{1}{1 + e^{-(\beta'X)}}$, \hspace{0.1cm} $\beta'X = \beta_{0} + \beta_{1}x_{1} + \cdots + \beta_{k}x_{k}$ \hspace{0.1cm} (1)\\
\end{center}

Based on the \textit{consensus} of our experts, we are able to specify a degree of future digitalization for 47 occupations. More than 75\% of our experts agreed that the characteristic tasks of these professions will change to a significant degree with the development of digital technologies and mobile robotics. With the use of the PIAAC data set, we are able to relate the degree of digitalization to personal characteristics and occupation-specific tasks. Based on this relationship, we estimate the degree of digitalization for all professions in the data set. In contrast to the work by \cite{frey2017future}, we apply local experts' opinions and perform our estimations on the basis of individual characteristics.\\

\section{Results}\label{sec:results}

\begin{figure}[h!]
\begin{center}
\includegraphics[width=0.9\linewidth, keepaspectratio, ]{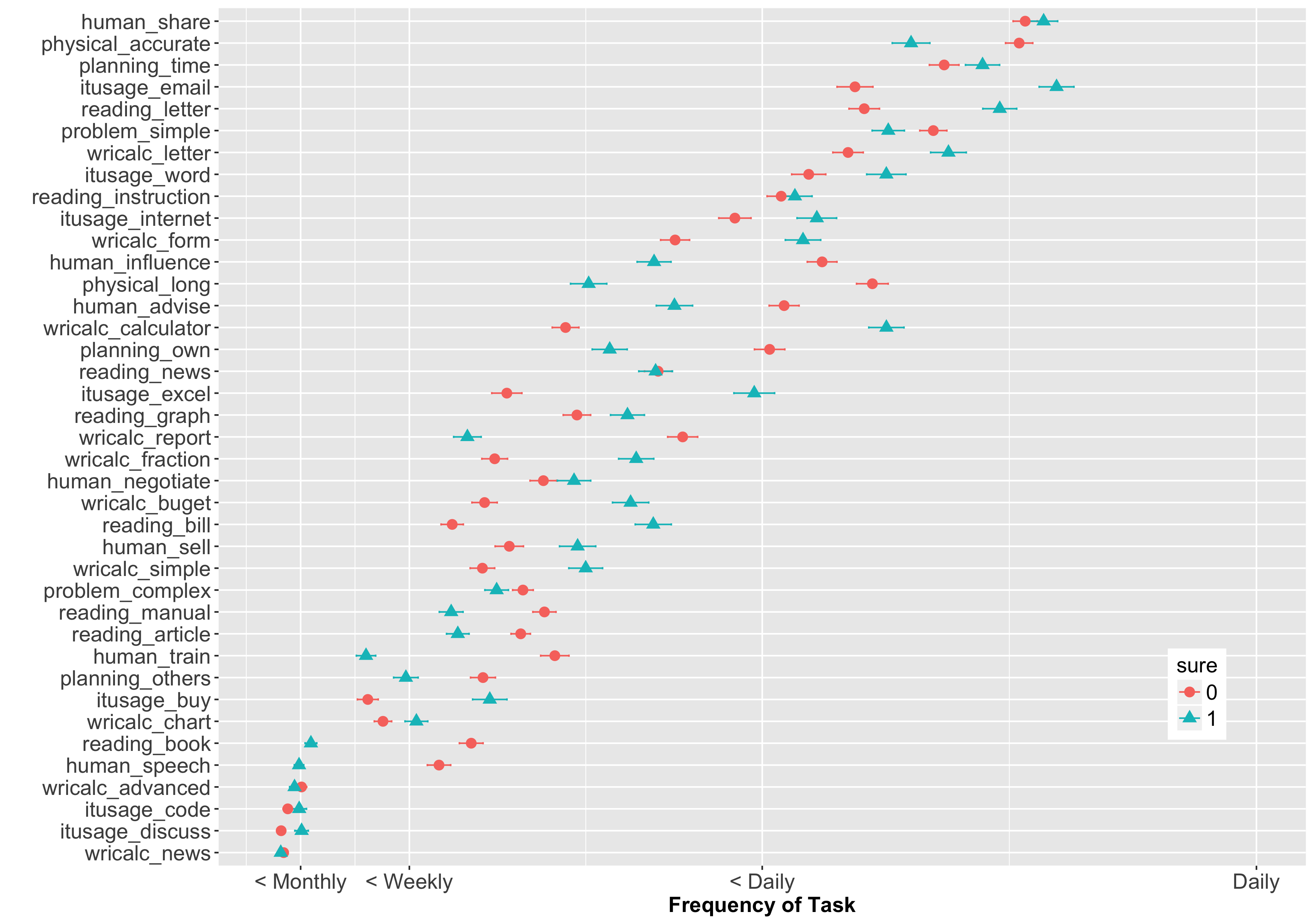}
\caption{\emph{Tasks at work}: Of the 39 tasks, some are performed on a daily basis, while others are carried out only once a month. Some of the activities vary significantly between jobs with a high and low degree of future digitalization.}
\label{fig:tasks01}
\end{center}
\end{figure}

For some tasks we see a clear relationship with the \textit{consensus} of our experts. In Figure \ref{fig:tasks01}, the frequencies of the 39 tasks are compared to the \textit{consensus} of our experts. On average, some tasks, such as coding (\textit{itusage\_code}), are, on average, performed less than once a month, while others, such as sharing information with others (\textit{human\_share}), are carried out on an almost daily basis. For some activities, prevalence does not differ significantly between the two \textit{consensus} job groups, for example, \textit{itusage\_code} or \textit{human\_share}. However, for most of the activities, a clear separation between the \textit{consensus} groups is visible. Activity involving long physical work (\textit{physical\_long}) is less commonly performed in professions that are expected to change during digitalization, according to our experts. Other activities show the exact opposite pattern. Calculating (\textit{wricalc\_calculator}) or the use of computer software \textit{Excel} (\textit{itusage\_excel}), for example, is much more prevalent in professions that are expected to change. This observation, confirmed by the findings of the inferential model, is a first indication that professions with a high degree of computer-based office routines are more likely to change in light of digital technologies.

\begin{center}
FIGURE \ref{fig:tasks01} ABOUT HERE
\end{center}

In addition to the 39 tasks, individual-, job- and firm-specific characteristics can help explain the consensus opinions of our experts, as shown in Table \ref{tab:modelcompare}. The final and full model (6) indicates that, apart from work activities, education, firm sector, job responsibility and training are related to the degree of future digitalization. Individuals with a high level of education, who work in a job that requires training or responsibility, are typically less likely to be employed in an occupation that is going to change significantly. Interestingly, our results indicate a non-linear relationship with education. Individuals with a medium level of schooling are employed in jobs with a higher level of future digitalization than workers with high or low levels of education. \\

Our model indicates that certain work activities are strongly related to the degree of digital change in the workplace. Tasks such as extracting complex information by reading books (\textit{reading\_book}) or writing non-routine content (\textit{wricalc\_report}) are related to professions with a low degree of technological change. On the other hand, activities such as calculations (\textit{wricalc\_calculator}) or extracting simple information (\textit{wricalc\_news}) are associated with a stronger change in the job profile in the near future. \cite{mcafee2014second}, for example, show that news stations have begun implementing algorithms that are able to write simple pieces in the context of sports or weather forecasts. Moreover, for professions that predominantly rely on physical labour, impacts of technological change are also low.

\begin{center}
FIGURE \ref{fig:distribution} ABOUT HERE
\end{center}

\begin{figure}[h!]
\begin{center}
\includegraphics[width=0.9\linewidth, keepaspectratio, ]{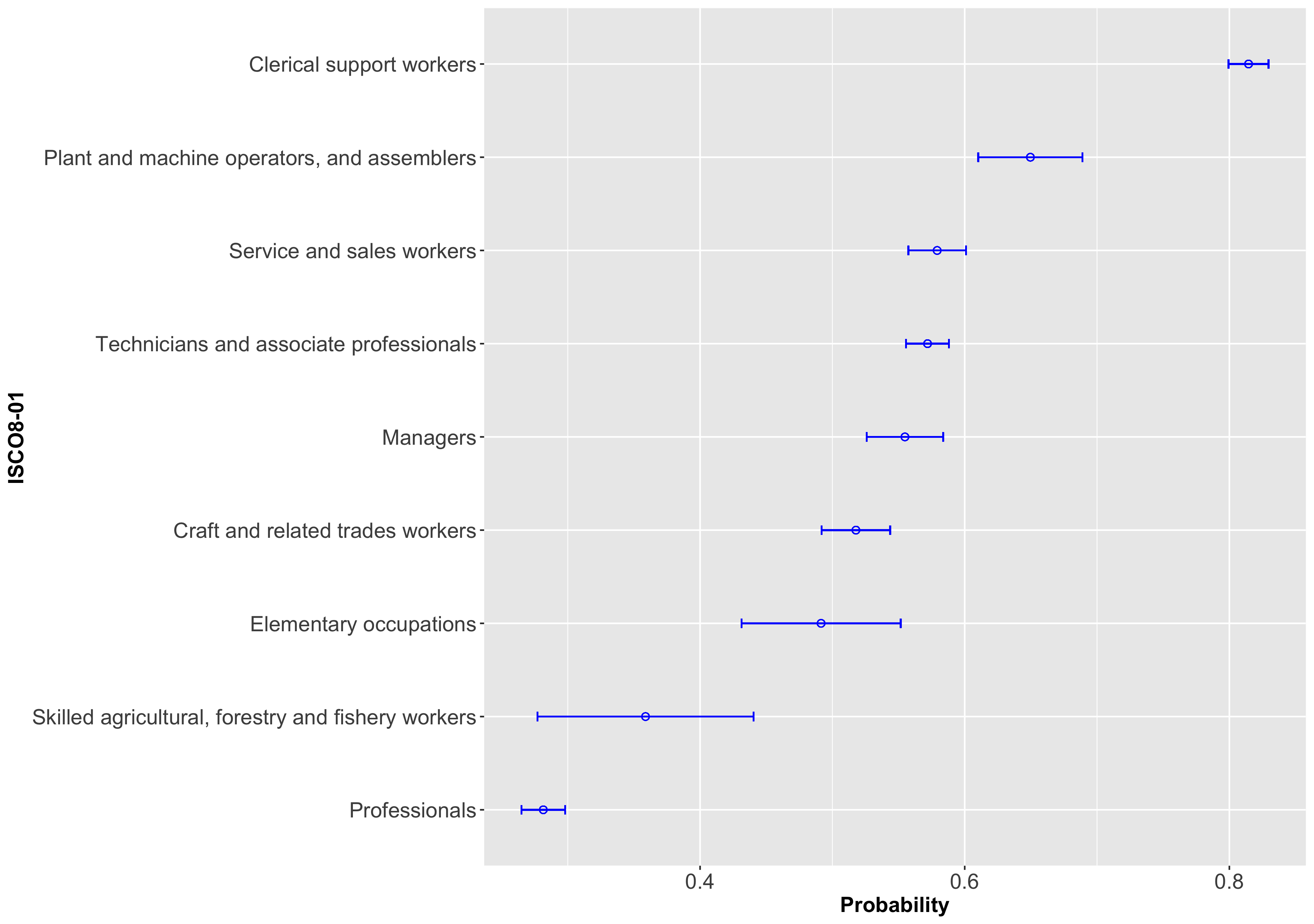}
\caption{\emph{ISCO Level 1}: For the top level of occupations, clerical professions have, by far, the highest risk of future digitalization. Professionals are at the lower end of the scale.}
\label{fig:isco_level1}
\end{center}
\end{figure}

\begin{figure}[h!]
\begin{center}
\includegraphics[width=0.9\linewidth, keepaspectratio, ]{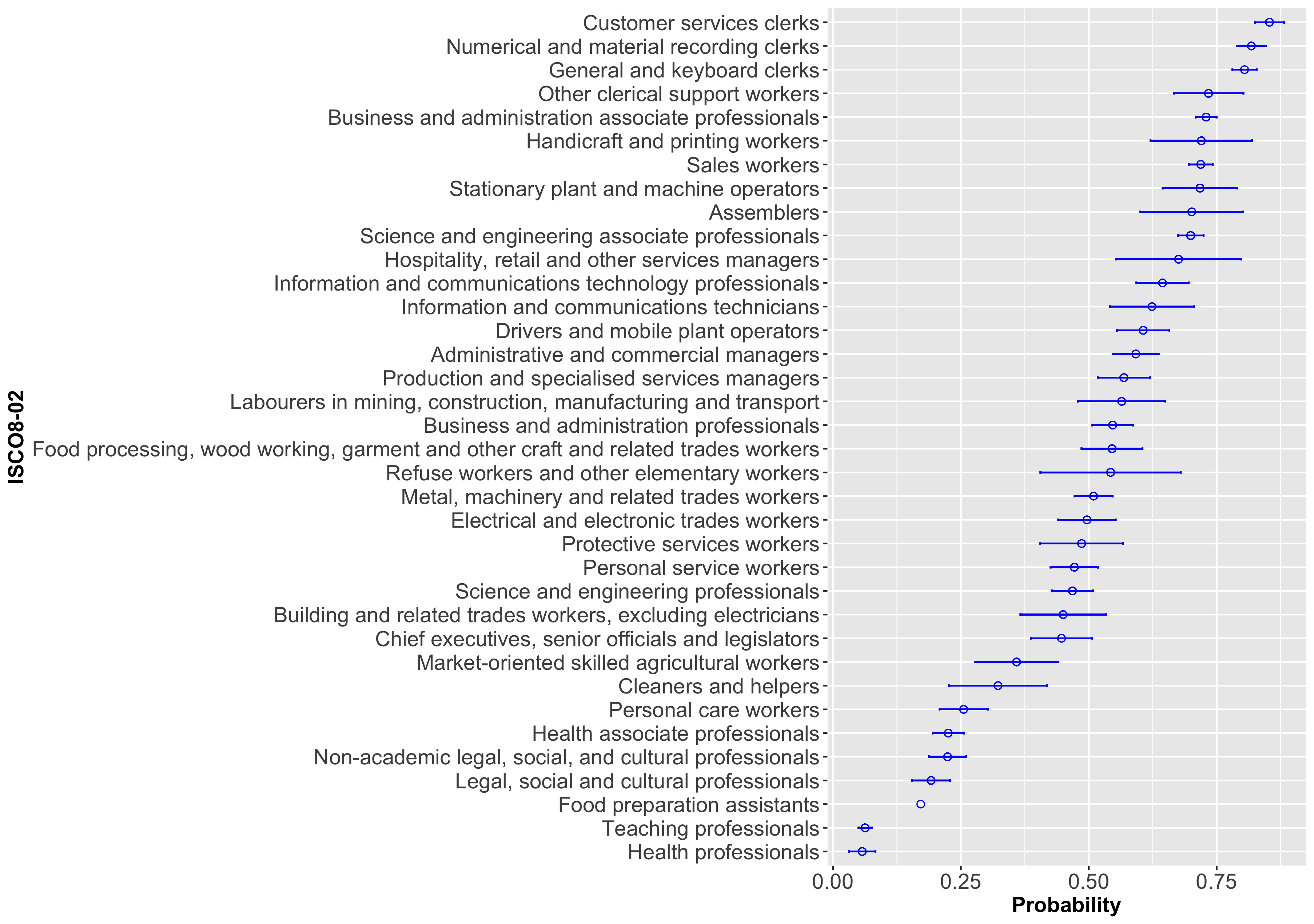}
\caption{\emph{ISCO Level 2}: Professional occupations involving teaching and healthcare, within legal, social or cultural environments, exhibit particularly low probabilities of digital transformation.}
\label{fig:isco_level2}
\end{center}
\end{figure}

Among occupations, there is a clear trend (Figures \ref{fig:isco_level1} and \ref{fig:isco_level2}): clerical support workers, who perform simple computer-based office routines, are highly susceptible to technological changes. This is in line with previous findings. On the other hand, professionals, who work with complex and unstructured information, and skilled workers in agricultural fields, who perform physical work, are less likely to experience major changes in their job profile. Professional occupations involving teaching and healthcare within legal, social or cultural environments (Figure \ref{fig:isco_level2}) exhibit particularly low probabilities of digital transformation. This finding is consistent for individuals working in a job that requires an academic degree, as well as for those without such a qualification. On average, most occupations show a probability of change between 40\% and 60\%. 

\begin{center}
FIGURES \ref{fig:isco_level1} and \ref{fig:isco_level2} ABOUT HERE
\end{center}

When comparing our model findings, clear differences emerge with regard to the degree of susceptibility in employment to digital technologies. However, our model testing suggests that these difference are mainly driven by model selection, rather than heterogeneity among tasks within occupations. Table \ref{tab:comparison} compares the set-ups of our research and previous studies.

\begin{center}
TABLE \ref{tab:comparison} ABOUT HERE
\end{center}

\begin{figure}[h!]
\begin{center}
\includegraphics[width=0.9\linewidth, keepaspectratio, ]{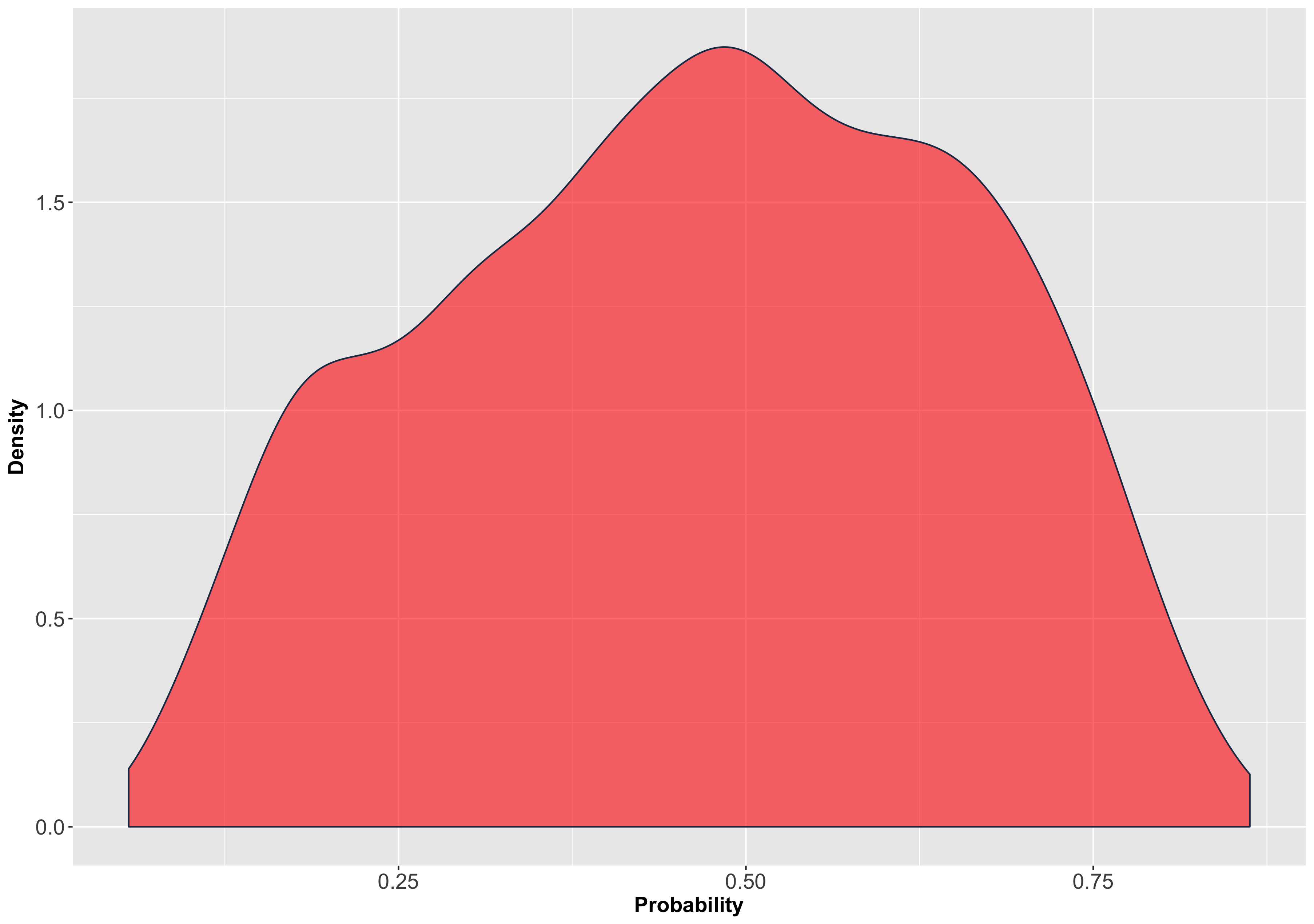}
\caption{\emph{Fractional model}: Similar to the work by \cite{arntz2016risk} and \cite{nagl2017digitalisierung}, the application of a fractional model \citep{papke1993econometric} yields a bell-shaped distribution of predicted probabilities.}
\label{fig:distribution_frac}
\end{center}
\end{figure}

\begin{figure}[h!]
\begin{center}
\includegraphics[width=0.9\linewidth, keepaspectratio, ]{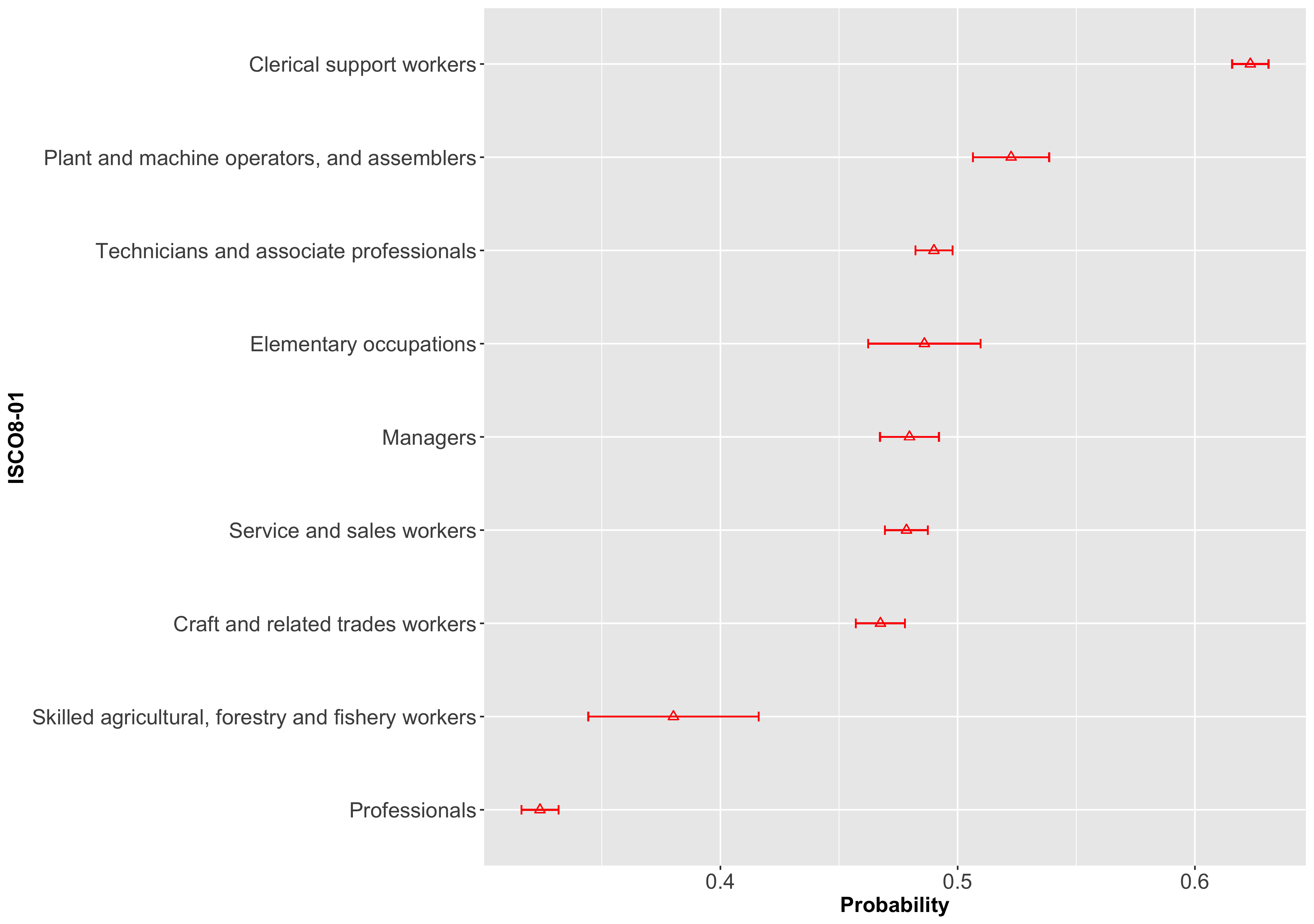}
\caption{\emph{ISCO Level 1}: The ranking of occupational classes does not change for the fractional model. However, predicted probabilities converge to the mean.}
\label{fig:isco_level1_frac}
\end{center}
\end{figure}

Two types of model settings are prevalent. \cite{frey2017future} start with binary opinions of experts and extrapolate them via a classification model for all occupations. \cite{bowles2014computerisation} directly transfers these estimations to European labour markets. Both studies conclude that a high share of workers (47\% in the US and 54\% in Austria) share a high risk of computerization. \cite{arntz2016risk} and \cite{nagl2017digitalisierung}, on the other hand, begin with discrete probabilities and apply a fractional model in order to extrapolate. In comparison, they show that only about 12\% and 9\%, respectively, have an automation risk of more than 70\%. In light of these contradictory findings, our model testing suggests that the different estimations are mainly due to the choice of model. Binary models yield a bimodal distribution of predicted probabilities with large high-risk groups. Fractional models lead to a bell-shaped distribution of probabilities with relatively low levels of high-risk individuals. Our own estimations for a fractional model (Figure \ref{fig:distribution_frac}) confirm this assumption. The ranking of occupational classes does not change significantly after the fractional model (Figure \ref{fig:isco_level1_frac}) has been used. However, predicted probabilities converge towards the mean.

\begin{center}
FIGURES \ref{fig:distribution_frac} AND \ref{fig:isco_level1_frac} ABOUT HERE
\end{center}

\begin{figure}[h!]
\begin{center}
\includegraphics[width=0.73\linewidth, keepaspectratio, ]{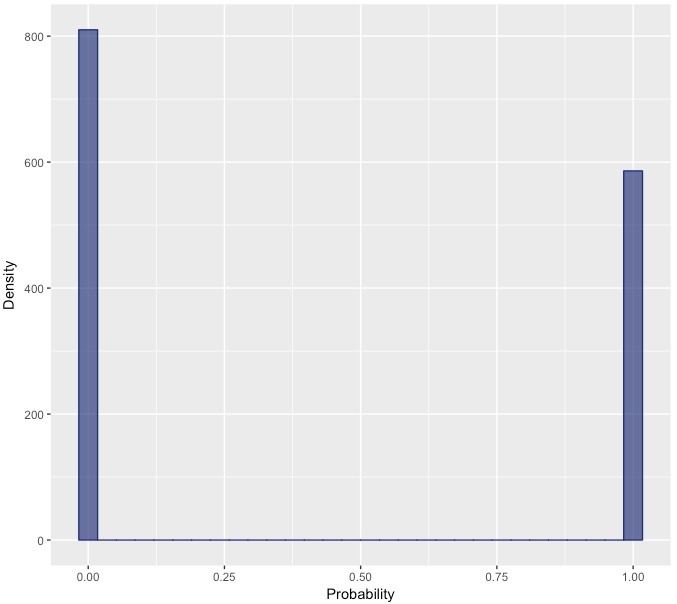}
\caption{\emph{Initial binary distribution}: Initially, in the logit model, slightly more individuals are labelled with a \textit{consensus} outcome of 0.}
\label{fig:binary_original}
\end{center}
\end{figure}

\begin{figure}[h!]
\begin{center}
\includegraphics[width=0.73\linewidth, keepaspectratio, ]{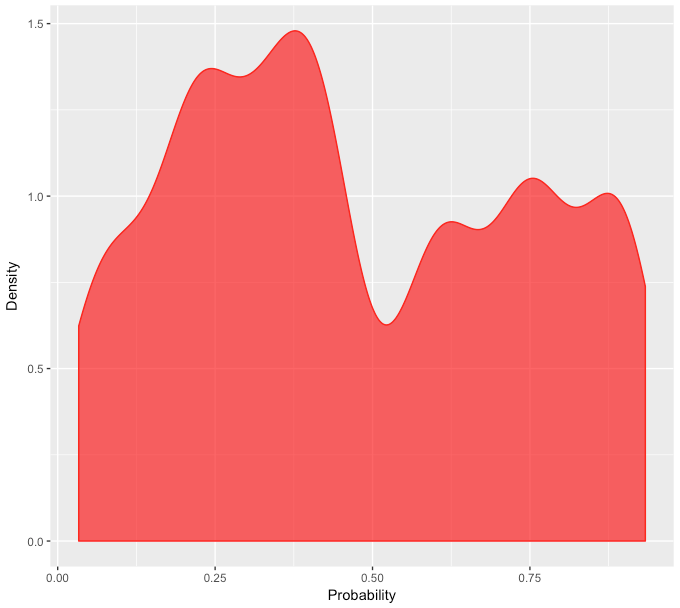}
\caption{\emph{Initial fractional distribution}: The initial distribution of the dependent variable in the fractional model is slightly skewed towards the lower values.}
\label{fig:fractional_original}
\end{center}
\end{figure}

When comparing the outcome of the binary and fractional model, the results of the latter contain a lower number of covariates, which are statistically relevant to the degree of digitalization. The fractional model, however, does not show any statistical significance concerning the covariates that have not been relevant in the binary model. In the fractional model, education and job responsibility show no statistical significance. Likewise, the tasks of speaking in front of humans, reading books, using words, and coding are not significant in the case of the fractional model environment. This general observation is not surprising from a statistical point of view, since the formally strict binary outcome in a small sample has now been changed to a smooth continuous scale in a sample twice the original size. However, it becomes clear that some covariates, such as physical work, writing reports, performing calculations or firm characteristics, are still aligned with the distribution of the fractional model. The distribution of other covariates has been polarized by the truncation of the binary model. The unconditional distributions of the binary and fractional models are shown in Figures \ref{fig:binary_original} and \ref{fig:fractional_original}.

\begin{center}
FIGURES \ref{fig:binary_original} AND \ref{fig:fractional_original} ABOUT HERE
\end{center}

Similarly, when moving the threshold of consensus from our final value towards 0.5, the outcome of the binary model starts to slightly approach the results of the fractional model. However, no significant changes appear, except for a deterioration in statistical significance.\\

\section{Conclusion}\label{sec:conclusion}

Our model explicitly diverges from the approach taken in previous contributions to this field. We assume that the diversity of previous estimations of job susceptibility stems from model specification. In order to test this assumption we conducted a case study with local expert opinions about near-term changes in occupations in Austria. This is a significant conceptual improvement in contrast to prior investigations \citep{arntz2016risk, bowles2014computerisation}, which studies rely on the judgement of machine learning experts concerning the US labour market, stemming from the workshop organized by \cite{frey2017future}. However, the authors do not allow for heterogeneity within the same profession. This limitation is ruled out by our model approach. Past findings are, in part, contradictory. 47\% of jobs in the US (54\% in Austria) share a high risk of automation, according to \cite{frey2017future} and \cite{bowles2014computerisation}, while \cite{arntz2016risk} and \cite{nagl2017digitalisierung} estimate this share to be 12\% and 9\%, respectively, for Austria. Our findings show that these differences are mainly driven by the selection of the model, and not so much by controlling for personal characteristics or tasks.\\

Our findings show that the tasks that humans perform during their typical working day are of significant importance when determining the impact of digital technologies on the future workspace. Activities such as extracting complex information by reading books or writing non-routine content reduce the impact of technologies. On the other hand, tasks such as calculations or extracting simple information will lead to a stronger change in job profiles in the next decade. Furthermore, as the current generation of technological progress has a stronger impact on cognitive and routine tasks than on physical labour, the extent of physical work within a job profile reduces the effect of digital change. Although the future of work will most likely be a complementary partnership between humans and computers, workers performing computer-related routine activities, such as spreadsheet calculations or Internet usage \citep{stephany2021does, stephany2021one, stephany2020coding}, are under stronger pressure to adapt. Our findings about the "inverse U-shaped" relationship between education and digitalization support previous hypotheses about the skill-based polarization of the labour market \citep{goos2009job}. This suggests further polarization in the near future.\\

Our results indicate that some jobs can expect to change more than others during the current phase of digital progress. This is surely not the first time in history that this has happened. During the Industrial Revolution, technological advancements made manufacturing jobs less intensive in terms of monotonous physical labour. In contrast to the age of the steam engine, today's technologies, such as algorithms, unfold their potential in disciplines that require routine cognitive effort. Typical computer-backed office tasks, such as in the clerical professions, are more exposed to digital transformation than occupations marked by physical labour. Likewise, jobs in which complex information is processed and that require a high level of education and training are less prone to digital change in the near future. Teaching and health-care professionals working within in legal, social or cultural environments belong to occupations with the lowest level of technological pressure. In the near future, these disciplines can be regarded as a sustainable choice for future generations seeking job security in unsteady times.\\ 

In addition, while most research focuses on human labour that can be replaced by technology, little attention has been given to the effect that digital technologies have on job creation. As our findings improve the understanding of the displacement effect of technologies, more research should be conducted in order to incorporate the effect of job creation, and in turn appreciate the full impact of the technological change on the labour market. 

\newpage
\begin{footnotesize}
\bibliographystyle{elsarticle-harv}

\bibliography{automation_lib}

\begin{thebibliography}{35}
\expandafter\ifx\csname natexlab\endcsname\relax\def\natexlab#1{#1}\fi
\expandafter\ifx\csname url\endcsname\relax
  \def\url#1{\texttt{#1}}\fi
\expandafter\ifx\csname urlprefix\endcsname\relax\def\urlprefix{URL }\fi

\bibitem[{Acemo\u{g}lu and Restrepo(2017)}]{acemoglu2017robots}
Acemo\u{g}lu, D., Restrepo, P., 2017. Robots and jobs: Evidence from us labor
  markets.

\bibitem[{Aghion and Howitt(1994)}]{aghion1994growth}
Aghion, P., Howitt, P., 1994. Growth and unemployment. The Review of Economic
  Studies 61~(3), 477--494.

\bibitem[{Arntz et~al.(2016)Arntz, Gregory, and Zierahn}]{arntz2016risk}
Arntz, M., Gregory, T., Zierahn, U., 2016. The risk of automation for jobs in
  oecd countries: A comparative analysis. OECD Social, Employment, and
  Migration Working Papers~(189).

\bibitem[{Autor(2013)}]{david2013task}
Autor, D., 2013. The" task approach" to labor markets: an overview. Tech. rep.,
  National Bureau of Economic Research.

\bibitem[{Autor(2015)}]{david2015there}
Autor, D., 2015. Why are there still so many jobs? the history and future of
  workplace automation. Journal of Economic Perspectives 29~(3), 3--30.

\bibitem[{Autor and Dorn(2013)}]{david2013growth}
Autor, D., Dorn, D., 2013. The growth of low-skill service jobs and the
  polarization of the us labor market. American Economic Review 103~(5),
  1553--97.

\bibitem[{Autor et~al.(2003)Autor, Levy, and Murnane}]{autor2003skill}
Autor, D., Levy, F., Murnane, R.~J., 2003. The skill content of recent
  technological change: An empirical exploration. The Quarterly Journal of
  Economics 118~(4), 1279--1333.

\bibitem[{Beaudry et~al.(2016)Beaudry, Green, and Sand}]{beaudry2016great}
Beaudry, P., Green, D.~A., Sand, B.~M., 2016. The great reversal in the demand
  for skill and cognitive tasks. Journal of Labor Economics 34~(S1),
  S199--S247.

\bibitem[{Bonin et~al.(2015)Bonin, Gregory, and Zierahn}]{bonin2015ubertragung}
Bonin, H., Gregory, T., Zierahn, U., 2015. {\"U}bertragung der studie von
  frey/osborne (2013) auf deutschland. Tech. rep., ZEW Kurzexpertise.

\bibitem[{Bowles(2014)}]{bowles2014computerisation}
Bowles, J., 2014. The computerisation of european jobs--who will win and who
  will lose from the impact of new technology onto old areas of employment.
  Bruegel Blog 17.

\bibitem[{Brandes and Wattenhofer(2016)}]{brandes2016opening}
Brandes, P., Wattenhofer, R., 2016. Opening the frey/osborne black box: Which
  tasks of a job are susceptible to computerization? arXiv preprint
  arXiv:1604.08823.

\bibitem[{Bresnahan(1999)}]{bresnahan1999computerisation}
Bresnahan, T.~F., 1999. Computerisation and wage dispersion: an analytical
  reinterpretation. The Economic Journal 109~(456), 390--415.

\bibitem[{Charles et~al.(2013)Charles, Hurst, and
  Notowidigdo}]{charles2013manufacturing}
Charles, K.~K., Hurst, E., Notowidigdo, M., 2013. Manufacturing decline,
  housing booms, and non-employment.

\bibitem[{Dengler and Matthes(2015)}]{dengler2015folgen}
Dengler, K., Matthes, B., 2015. Folgen der digitalisierung f{\"u}r die
  arbeitswelt: Substituierbarkeitspotenziale von berufen in deutschland. Tech.
  rep., IAB-Forschungsbericht.

\bibitem[{Dustmann et~al.(2009)Dustmann, Ludsteck, and
  Sch{\"o}nberg}]{dustmann2009revisiting}
Dustmann, C., Ludsteck, J., Sch{\"o}nberg, U., 2009. Revisiting the german wage
  structure. The Quarterly Journal of Economics 124~(2), 843--881.

\bibitem[{Ford(2015)}]{ford2015rise}
Ford, M., 2015. Rise of the Robots: Technology and the Threat of a Jobless
  Future. Basic Books.

\bibitem[{Frey and Osborne(2017)}]{frey2017future}
Frey, C.~B., Osborne, M., 2017. The future of employment. How susceptible are
  jobs to computerisation.

\bibitem[{Goldin and Katz(1998)}]{goldin1998origins}
Goldin, C., Katz, L.~F., 1998. The origins of technology-skill complementarity.
  The Quarterly Journal of Economics 113~(3), 693--732.

\bibitem[{Goos et~al.(2009)Goos, Manning, and Salomons}]{goos2009job}
Goos, M., Manning, A., Salomons, A., 2009. Job polarization in europe. American
  economic review 99~(2), 58--63.

\bibitem[{Hoffmann et~al.(2019)Hoffmann, Dimitrova, Muttarak, Crespo~Cuaresma,
  Stephany, and Peisker}]{hoffmann2019quantifying}
Hoffmann, R., Dimitrova, A., Muttarak, R., Crespo~Cuaresma, J., Stephany, F.,
  Peisker, J., 2019. Quantifying the evidence on environmental migration: A
  meta-analysis on country-level studies. In: Population Association of America
  Annual Meeting, Austin, USA.

\bibitem[{Jaimovich and Siu(2012)}]{jaimovich2012trend}
Jaimovich, N., Siu, H.~E., 2012. The trend is the cycle: Job polarization and
  jobless recoveries. Tech. rep., National Bureau of Economic Research.

\bibitem[{James et~al.(2013)James, Witten, Hastie, and
  Tibshirani}]{james2013introduction}
James, G., Witten, D., Hastie, T., Tibshirani, R., 2013. An introduction to
  statistical learning. Vol. 112. Springer.

\bibitem[{McAfee and Brynjolfsson(2014)}]{mcafee2014second}
McAfee, A., Brynjolfsson, E., 2014. The second machine age. WW Norton.

\bibitem[{Michaels et~al.(2014)Michaels, Natraj, and
  Van~Reenen}]{michaels2014has}
Michaels, G., Natraj, A., Van~Reenen, J., 2014. Has ict polarized skill demand?
  evidence from eleven countries over twenty-five years. Review of Economics
  and Statistics 96~(1), 60--77.

\bibitem[{Nagl et~al.(2017)Nagl, Titelbach, and
  Valkova}]{nagl2017digitalisierung}
Nagl, W., Titelbach, G., Valkova, K., 2017. Digitalisierung der arbeit:
  Substituierbarkeit von berufen im zuge der automatisierung durch industrie
  4.0; endbericht.

\bibitem[{Pajarinen et~al.(2014)Pajarinen, Rouvinen,
  et~al.}]{pajarinen2014computerization}
Pajarinen, M., Rouvinen, P., et~al., 2014. Computerization threatens one third
  of finnish employment. ETLA Brief 22~(13.1), 2014.

\bibitem[{Papke and Wooldridge(1993)}]{papke1993econometric}
Papke, L.~E., Wooldridge, J., 1993. Econometric methods for fractional response
  variables with an application to 401 (k) plan participation rates.

\bibitem[{Peneder et~al.(2016)Peneder, Bock-Schappelwein, Firgo, Fritz, and
  Streicher}]{peneder2016osterreich}
Peneder, M., Bock-Schappelwein, J., Firgo, M., Fritz, O., Streicher, G., 2016.
  {\"O}sterreich im Wandel der Digitalisierung. WIFO, {\"O}sterreichisches
  Institut f{\"u}r Wirtschaftsforschung.

\bibitem[{Schumpeter(1942)}]{schumpeter1942creative}
Schumpeter, J., 1942. Creative destruction. Capitalism, socialism and democracy
  825.

\bibitem[{Spitz-Oener(2006)}]{spitz2006technical}
Spitz-Oener, A., 2006. Technical change, job tasks, and rising educational
  demands: Looking outside the wage structure. Journal of labor economics
  24~(2), 235--270.

\bibitem[{Stephany(2017)}]{stephany2017your}
Stephany, F., 2017. Who are your joneses? socio-specific income inequality and
  trust. Social indicators research 134~(3), 877--898.

\bibitem[{Stephany(2019)}]{stephany2019deepens}
Stephany, F., 2019. It deepens like a coastal shelf: Educational mobility and
  social capital in germany. Social Indicators Research 142~(2), 855--885.

\bibitem[{Stephany(2021{\natexlab{a}})}]{stephany2021one}
Stephany, F., 2021{\natexlab{a}}. One size does not fit all: Constructing
  complementary digital reskilling strategies using online labour market data.
  Big Data \& Society 8~(1), 20539517211003120.

\bibitem[{Stephany(2021{\natexlab{b}})}]{stephany2021does}
Stephany, F., 2021{\natexlab{b}}. When does it pay off to learn a new skill?
  revealing the complementary benefit of cross-skilling.

\bibitem[{Stephany et~al.(2020)Stephany, Braesemann, and
  Graham}]{stephany2020coding}
Stephany, F., Braesemann, F., Graham, M., 2020. Coding together--coding alone:
  the role of trust in collaborative programming. Information, Communication \&
  Society, 1--18.

\end{thebibliography}
\end{footnotesize}

\pagebreak

\newpage

\newpage

\newpage
 
\newpage

\newpage
\begin{table}[h!] \centering 
\centering
\caption{Our experts gave their yes/no responses in relation to the 100 most common professions in Austria. The \textit{mean}, \textit{mode} and \textit{consensus} (at least 75\% responded with yes or no) were calculated for each profession.}
\label{tab:top100}
\scalebox{0.8}{ 
\begin{tabular}{llccc}
\\[-1.8ex]\hline 
\hline \\[-1.8ex] 
\textbf{ISCO-08-4 Code} & \textbf{ISCO-08-4 Name}                                                                     & \textbf{Mean}  & \textbf{Mode} & \textbf{\textit{Consensus} }\\
\\[-1.8ex]\hline 
5311          & Childcare workers                                                                & 0.033 & 0    & 0    \\
5120          & Cooks                                                                             & 0.034 & 0    & 0    \\
3255          & Physiotherapy technicians and assistants                                          & 0.038 & 0    & 0    \\
2652          & Musicians, singers and composers                                                  & 0.067 & 0    & 0    \\
5141          & Hairdressers                                                                      & 0.067 & 0    & 0    \\
3412          & Social work associate professionals                                               & 0.071 & 0    & 0    \\
5412          & Police officers                                                                   & 0.071 & 0    & 0    \\
2341          & Primary school teachers                                                           & 0.074 & 0    & 0    \\
2635          & Social work and counselling professionals                                         & 0.103 & 0    & 0    \\
3355          & Police inspectors and detectives                                                  & 0.103 & 0    & 0    \\
5321          & Healthcare assistants                                                            & 0.103 & 0    & 0    \\
6113          & Gardeners, horticultural and nursery growers                                      & 0.111 & 0    & 0    \\
2161          & Building architects                                                               & 0.148 & 0    & 0    \\
6130          & Mixed crop and animal producers                                                   & 0.148 & 0    & 0    \\
7421          & Electronics mechanics and servicers                                               & 0.160 & 0    & 0    \\
2212          & Specialist medical practitioners                                                  & 0.172 & 0    & 0    \\
2310          & University and higher education teachers                                          & 0.172 & 0    & 0    \\
5131          & Waiters                                                                           & 0.179 & 0    & 0    \\
7126          & Plumbers and pipe fitters                                                         & 0.179 & 0    & 0    \\
7512          & Bakers, pastry cooks and confectionery makers                                     & 0.185 & 0    & 0    \\
1349          & Professional services managers not elsewhere classified                           & 0.192 & 0    & 0    \\
7412          & Electrical mechanics and fitters                                                  & 0.200 & 0    & 0    \\
1323          & Construction managers                                                             & 0.207 & 0    & 0    \\
1411          & Hotel managers                                                                    & 0.207 & 0    & 0    \\
3221          & Nursing associate professionals                                                   & 0.222 & 0    & 0    \\
2330          & Secondary education teachers                                                      & 0.231 & 0    & 0    \\
7411          & Building and related electricians                                                 & 0.240 & 0    & 0    \\
3259          & Health associate professionals not elsewhere classified                           & 0.250 & 0    & 0    \\
5151          & Cleaning/housekeeping supervisors in offices, hotels and others 				  & 0.250 & 0    & 0    \\
2142          & Civil engineers                                                                   & 0.259 & 0    & .    \\
2149          & Engineering professionals not elsewhere classified                                & 0.261 & 0    & .    \\
2642          & Journalists                                                                       & 0.267 & 0    & .    \\
1321          & Manufacturing managers                                                            & 0.286 & 0    & .    \\
2611          & Lawyers                                                                           & 0.296 & 0    & .    \\
2359          & Teaching professionals not elsewhere classified                                   & 0.304 & 0    & .    \\
2144          & Mechanical engineers                                                              & 0.308 & 0    & .    \\
3251          & Dental assistants and therapists                                                  & 0.321 & 0    & .    \\
3411          & Police inspectors and detectives                                                  & 0.321 & 0    & .    \\
3256          & Medical assistants                                                                & 0.333 & 0    & .    \\
2166          & Graphic and multimedia designers                                                  & 0.345 & 0    & .    \\
2631          & Economists                                                                        & 0.346 & 0    & .    \\
7233          & Agricultural and industrial machinery mechanics and repairers                     & 0,346 & 0    & .    \\
7522          & Cabinet makers and related workers                                                & 0.348 & 0    & .    \\
2512          & Software developers                                                               & 0.357 & 0    & .    \\
5414          & Security guards                                                                   & 0.357 & 0    & .    \\
9112          & Cleaners and helpers in offices, hotels and other establishments                  & 0.357 & 0    & .    \\
7119          & Building frame and related trades workers not elsewhere classified                & 0.360 & 0    & .    \\
2421          & Management and organization analysts                                              & 0.370 & 0    & .    \\
5153          & Building caretakers                                                               & 0.370 & 0    & .    \\
7112          & Bricklayers and related workers                                                   & 0.370 & 0    & .    \\
9412          & Kitchen helpers                                                                   & 0.385 & 0    & .    \\
3257          & Environmental and occupational health inspectors and associates                   & 0.400 & 0    & .    \\
7231          & Motor vehicle mechanics and repairers                                             & 0.400 & 0    & .    \\
2431          & Advertising and marketing professionals                                           & 0.407 & 0    & .    \\
7212          & Welders and flame cutters                                                          & 0.407 & 0    & .    \\
1324          & Supply, distribution and related managers                                         & 0.414 & 0    & .    \\
3359          & Regulatory government associate professionals						              & 0.423 & 0    & .    \\
5223          & Shop sales assistants                                                             & 0.423 & 0    & .    \\
7543          & Product graders and testers (excluding foods and beverages)                       & 0.458 & 0    & .    \\
3115          & Mechanical engineering technicians                                                & 0.462 & 0    & .    \\
2262          & Pharmacists                                                                       & 0.500 & 0    & .    \\
\hline
\hline \\[-1.8ex] 
\end{tabular}}
\end{table}

\newpage
\begin{table}[h!] \centering 
\centering
\scalebox{0.8}{ 
\begin{tabular}{llccc}
\\[-1.8ex]\hline 
\hline \\[-1.8ex] 
\textbf{ISCO-08-4 Code} & \textbf{ISCO-08-4 Name}                                                                     & \textbf{Mean}  & \textbf{Mode} & \textbf{\textit{Consensus} }\\
\\[-1.8ex]\hline 
9629          & Elementary workers not elsewhere classified                                       & 0.500 & 0    & .    \\
7214          & Structural metal preparers and erectors                                           & 0.520 & 1    & .    \\
7523          & Woodworking machine tool setters and operators                                    & 0.520 & 1    & .    \\
8219          & Assemblers not elsewhere classified                                               & 0.538 & 1    & .    \\
3353          & Government social benefits officials                                              & 0.556 & 1    & .    \\
3352          & Government tax and excise officials                                               & 0.571 & 1    & .    \\
8212          & Electrical and electronic equipment assemblers                                    & 0.577 & 1    & .    \\
9332          & Drivers of animal-drawn vehicles and machinery                                    & 0.577 & 1    & .    \\
7223          & Metal working machine tool setters and operators                                  & 0.583 & 1    & .    \\
3322          & Commercial sales representatives                                                  & 0.593 & 1    & .    \\
3323          & Buyers                                                                            & 0.593 & 1    & .    \\
3334          & Real estate agents and property managers                                          & 0.607 & 1    & .    \\
4120          & Secretaries (general)                                                             & 0.607 & 1    & .    \\
2411          & Accountants                                                                       & 0.633 & 1    & .    \\
7321          & Pre-press technicians                                                             & 0.640 & 1    & .    \\
9329          & Manufacturing labourers not elsewhere classified                                  & 0.640 & 1    & .    \\
9333          & Freight handlers                                                                  & 0.652 & 1    & .    \\
8160          & Food and related products machine operators                                       & 0.667 & 1    & .    \\
9334          & Shelf fillers                                                                     & 0.692 & 1    & .    \\
9621          & Messengers, package deliverers and luggage porters                                & 0.692 & 1    & .    \\
3118          & Draughtspersons                                                                   & 0.720 & 1    & .    \\
3313          & Accounting associate professionals                                                & 0.731 & 1    & .    \\
4110          & General office clerks                                                             & 0.750 & 1    & 1    \\
8322          & Car, taxi and van drivers                                                         & 0.759 & 1    & 1    \\
8131          & Chemical products plant and machine operators                                     & 0.792 & 1    & 1    \\
8332          & Heavy truck and lorry drivers                                                     & 0.793 & 1    & 1    \\
8121          & Metal processing plant operators                                                  & 0.800 & 1    & 1    \\
8122          & Metal finishing, plating and coating machine operators                            & 0.800 & 1    & 1    \\
4321          & Stock clerks                                                                      & 0.828 & 1    & 1    \\
4412          & Mail carriers and sorting clerks                                                  & 0.862 & 1    & 1    \\
3324          & Trade brokers                                                                     & 0.867 & 1    & 1    \\
4322          & Production clerks                                                                 & 0.875 & 1    & 1    \\
4312          & Statistical, finance and insurance clerks                                         & 0.897 & 1    & 1    \\
5230          & Cashiers and ticket clerks                                                        & 0.897 & 1    & 1    \\
3321          & Insurance representatives                                                         & 0.900 & 1    & 1    \\
4222          & Contact centre information clerks                                                 & 0.900 & 1    & 1    \\
4323          & Transport clerks                                                                  & 0.926 & 1    & 1    \\
4311          & Accounting and bookkeeping clerks                                                 & 0.933 & 1    & 1   \\
\hline
\hline \\[-1.8ex] 
\end{tabular}}
\end{table}

\newpage
\begin{table}[h!] \centering 
\centering
\caption{The PIAAC survey asked employees in Austria and Germany about how frequently they undertake the following 39 tasks at work.}
\label{tab:tasklist}
\scalebox{0.7}{ 
\begin{tabular}{lll}
\\[-1.8ex]\hline 
\hline \\[-1.8ex] 
Task Description  & Task Group & PIAAC Code\\
\hline 
Sharing work-related information with co-workers                                                                     & Human Interaction         
							& human\_share	\\
Instructing, training or teaching people, individually or in groups                                                  & Human Interaction         
							& human\_train	\\
Making speeches or giving presentations in front of five or more people                                              & Human Interaction         
							& human\_speech	\\
Selling a product or selling a service                                                                               & Human Interaction        
							& human\_sell	\\
Advising people                                                                                                      & Human Interaction        
							& human\_advise	\\
Persuading or influencing people                                                                                     & Human Interaction         
							& human\_influence	\\
Negotiating with people either inside or outside one's firm or organization                                          & Human Interaction         
							& human\_negotiate	\\
Using email                                                                                                            & IT Usage                  
							& itusage\_email	\\
Using the Internet in order to better understand issues related to one's work                                          & IT Usage                  
							& itusage\_internet	\\
Conducting transactions over the Internet, e.g., buying or selling                 
			 & IT Usage                  
             				& itusage\_buy	\\
Using spreadsheet software, for example, Excel                                                                          & IT Usage                  
							& itusage\_excel	\\
Using a word-processing package, for example, Word                                                                              & IT Usage                  
							& itusage\_word	\\
Using a programming language to program or write computer code                                                        & IT Usage                  
							& itusage\_code	\\
Participating in real-time discussions over the Internet, e.g., online conferences                 
			 & IT Usage                  
             				& itusage\_discuss	\\
Working physically for a long period                                                                                 & Physical Work             
							& physical\_long	\\
Using skill or accuracy with hands or fingers                                                                        & Physical Work             
							& physical\_accurate	\\
Planning one's own activities                                                                                        & Planning                  
							& planning\_own	\\
Planning the activities of others                                                                                    & Planning                  
							& planning\_others	\\
Organizing one's own time                                                                                            & Planning                  
							& planning\_time	\\
Solving simple problems, which require no more than 5 min of attention                                             & Problem-solving           
							& problem\_simple	\\
Solving complex problems, which require at least 30 min of attention                                               & Problem-solving           
							& problem\_complex	\\
Reading directions or instructions                                                                                      & Reading and Understanding 
							& reading\_instruction	\\
Reading letters, memos or emails                                                                                       & Reading and Understanding 
							& reading\_letter	\\
Reading articles in newspapers, magazines or newsletters                                                                & Reading and Understanding 
							& reading\_news	\\
Reading articles in professional journals or scholarly publications                                                     & Reading and Understanding 
							& reading\_article	\\
Reading books                                                                                                           & Reading and Understanding 
							& reading\_book	\\
Reading manuals or reference materials                                                                                  & Reading and Understanding 
							& reading\_manual	\\
Reading bills, invoices, bank statements or other financial statements                                                  & Reading and Understanding 
							& reading\_bill	\\
Reading diagrams, maps or schematics                                                                                    & Reading and Understanding 
							& reading\_graph	\\
Writing letters, memos or emails                                                                                      & Writing and Calculating   
							& wricalc\_letter	\\
Writing articles for newspapers, magazines or newsletters                                                              & Writing and Calculating   
							& wricalc\_news	\\
Writing reports                                                                                                        & Writing and Calculating   
							& wricalc\_report	\\
Filling in forms                                                                                                        & Writing and Calculating   
							& wricalc\_form	\\
Calculating prices, costs or budgets                                                                                   & Writing and Calculating   
							& wricalc\_budget	\\
Using or calculating fractions, decimals or percentages                                                                  & Writing and Calculating   
							& wricalc\_fraction	\\
Using a calculator(either hand-held or computer- based)                                                               & Writing and Calculating   
							& wricalc\_calculator	\\
Preparing charts, graphs or tables                                                                                     & Writing and Calculating   
							& wricalc\_chart	\\
Using simple algebra or formulas                                                                                       & Writing and Calculating   
							& wricalc\_simple	\\
Using more advanced mathematics or statistics, such as calculus,  
				& Writing and Calculating  & wricalc\_advanced	\\
complex algebra, trigonometry, or using of regression techniques &   &
\\
\hline
\hline \\[-1.8ex] 
\end{tabular}}
\end{table}

\begin{table}[h!] \centering 
\centering
\caption{Summary of characteristics}
\label{tab:summary}
\scalebox{0.8}{ 
\begin{tabular}{lllllll}
\\[-1.8ex]\hline 
\hline \\[-1.8ex] 
Name & Observations &  & &  & &              \\
\hline 
Age Group &  \\                             
"<16-19" & 104 \\
"20-24" & 195\\
"25-29" & 255 \\ 
"30-34" & 255 \\ 
"35-39" & 273 \\ 
"40-44" & 309 \\
"45-49" & 267 \\ 
"50-54" & 242 \\
"55-59" & 119 \\ 
">60" & 32 \\
& \\
Gender & \\
Male & 1,002 \\   
Female & 1,049 \\
& \\
Firm - Sector &  \\
"Public or NGO" & 668 \\
"Private" & 1,383  \\
& \\
Firm - Size &  \\    
"1-10" & 458 \\ 
"11-50" & 607 \\ 
"51-250" & 480 \\ 
"251-1000" & 326 \\ 
">1000" & 180 \\
& \\
Job -  Responsibility & \\
Yes & 1,182\\
No & 869\\
& \\
Job - Experience & \\ 
"<1 month" & 585\\ 
"1 to 6 months" & 282\\ 
"7 to 11 months" & 157\\ 
"1 or 2 years" & 472\\ 
"3 years or more" & 555\\
& \\  
Job - Education & \\ 
"<ISCED 3" & 252\\       
"ISCED 3-4" & 1,169\\ 
"ISCED 5+" & 630\\
& \\
&  & Min. & 25\% & Mean & 75\% & Max.\\
Education & &  \\
"Years in Full-time Education" & & 4.00 & 13.00 & 14.29 & 16.00 & 20.00 \\
&& \\
Skills && \\
Problem-solving & &     168.1 &  268.1   &  290.1 &  313.6 & 404.3      \\
Numeracy & &       160.4  &  269.6  &    294.8  &  321.7 &   409.7 2     \\
Literacy & &  156.8&    263.6   &   285.8  &  310.5 &   396.2 \\
\hline
\hline \\[-1.8ex] 
\end{tabular}}
\end{table}

\newpage
\begin{table}[h!] \centering 
\caption{Models (1)-(6) work with a binary outcome of job digitalisation, while the outcome of model (7) is continuously measured between 0 and 1. The full binary model (6) with personal/firm-/job-level characteristics and tasks shows the best model fit.}
\label{tab:modelcompare}
\scalebox{0.8}{
\begin{tabular}{@{\extracolsep{5pt}}lccccccc} 
\tiny
\\[-1.8ex]\hline 
\hline \\[-1.8ex] 
 & \multicolumn{7}{c}{\textit{Dependent variable}} \\ 
\cline{2-8} 
\\[-1.8ex] & Binary & Binary & Binary & Binary & Binary & Binary & Fractional \\ 
\\[-1.8ex] & (1) & (2) & (3) & (4) & (5) & (6) & (7)\\ 
\hline \\[-1.8ex] 
    &  &  & &  &  &  &  \\ 
 \textbf{Personal}&  &  & &  &  &  &  \\  
 Age Group & 0.069$^{*}$ & 0.106$^{**}$ & 0.175$^{***}$ & 0.105 &  & 0.141 & 0.025 \\ 
  & (0.040) & (0.046) & (0.057) & (0.068) &  & (0.094) & (0.037) \\ 
  Gender & $-$0.021 & 0.466$^{**}$ & 0.565$^{**}$ & 0.130 &  & 0.183 & 0.051 \\ 
  \textit{(Ref. male)}& (0.192) & (0.225) & (0.253) & (0.337) &  & (0.419) & (0.166) \\ 
  Education & 0.952$^{***}$ & 0.723$^{**}$ & 1.005$^{***}$ & 1.411$^{***}$ &  & 1.489$^{***}$ & 0.133 \\ 
  \textit{(Years)} & (0.283) & (0.305) & (0.355) & (0.419) &  & (0.495) & (0.230) \\ 
  Education$^{2}$ & $-$0.050$^{***}$ & $-$0.038$^{***}$ & $-$0.049$^{***}$ & $-$0.071$^{***}$ &  & $-$0.070$^{***}$ & $-$0.008 \\ 
  & (0.011) & (0.012) & (0.014) & (0.017) &  & (0.020) & (0.008) \\ 
    &  &  & &  &  &  &  \\ 
 \textbf{Firm}&  &  & &  &  &  &  \\  
  Firm - Sector &  & 2.300$^{***}$ & 2.170$^{***}$ &  &  & 1.695$^{***}$ & 0.522$^{***}$ \\ 
  \textit{(Ref. public)}&  & (0.240) & (0.263) &  &  & (0.454) & (0.188) \\ 
  Firm - Size &  & 0.085 & 0.129 &  &  & 0.408$^{**}$ & 0.104$^{*}$ \\ 
 \textit{(0-4)} &  & (0.090) & (0.100) &  &  & (0.168) & (0.061) \\ 
      &  &  & &  &  &  &  \\ 
 \textbf{Job}&  &  & &  &  &  &  \\  
  Job - Responsibility &  &  & $-$0.200 &  &  & $-$0.979$^{**}$ & $-$0.039 \\ 
  \textit{(Ref. no responsibility)}&  &  & (0.264) &  &  & (0.486) & (0.178) \\ 
  Job - Experience &  &  & 0.151$^{*}$ &  &  & $-$0.006 & $-$0.006 \\ 
  \textit{(0-4)}&  &  & (0.087) &  &  & (0.151) & (0.055) \\ 
  Job - Education &  &  & $-$1.293$^{***}$ &  &  & $-$1.919$^{***}$ & $-$0.483$^{***}$ \\ 
  \textit{(0-2)}&  &  & (0.230) &  &  & (0.424) & (0.148) \\ 
  Skill - Problem-solving &  &  & $-$0.0003 &  &  & $-$0.018$^{*}$ & $-$0.001 \\ 
  \textit{(Test Score)}&  &  & (0.006) &  &  & (0.010) & (0.004) \\ 
  Skill - Numeracy &  &  & 0.021$^{***}$ &  &  & 0.008 & 0.003 \\ 
  \textit{(Test score)}&  &  & (0.007) &  &  & (0.011) & (0.005) \\ 
  Skill - Literacy &  &  & $-$0.006 &  &  & 0.009 & $-$0.003 \\ 
  \textit{(Test score)}&  &  & (0.008) &  &  & (0.013) & (0.005) \\ 
  Cooperate with Humans &  &  & $-$0.360 &  &  & $-$0.307 & $-$0.167 \\ 
 \textit{(Frequency)}&  &  & (0.342) &  &  & (0.589) & (0.225) \\ 
  &  &  & &  &  &  &  \\ 
 \textbf{Tasks}&  &  & &  &  &  &  \\  
  human\_share &  &  &  & 0.769 & 0.730 & 0.880 & 0.083 \\ 
  &  &  &  & (0.507) & (0.480) & (0.619) & (0.254) \\ 
  human\_train &  &  &  & $-$0.263 & $-$0.483 & 0.183 & $-$0.090 \\ 
  &  &  &  & (0.550) & (0.522) & (0.688) & (0.249) \\ 
  human\_speech &  &  &  & $-$2.445$^{**}$ & $-$3.299$^{***}$ & $-$2.562$^{**}$ & $-$0.257 \\ 
  &  &  &  & (1.020) & (0.966) & (1.188) & (0.357) \\ 
  human\_sell &  &  &  & 0.765$^{*}$ & 0.847$^{**}$ & 0.966$^{*}$ & 0.117 \\ 
  &  &  &  & (0.455) & (0.411) & (0.550) & (0.214) \\ 
  human\_advise &  &  &  & $-$0.507 & $-$0.560 & $-$0.622 & $-$0.109 \\ 
  &  &  &  & (0.443) & (0.419) & (0.519) & (0.216) \\ 
  human\_influence &  &  &  & $-$1.236$^{***}$ & $-$1.081$^{**}$ & $-$0.886 & $-$0.249 \\ 
  &  &  &  & (0.474) & (0.429) & (0.559) & (0.212) \\ 
  human\_negotiate &  &  &  & 0.478 & 0.369 & 0.891 & 0.115 \\ 
  &  &  &  & (0.555) & (0.494) & (0.660) & (0.236) \\ 
  itusage\_email &  &  &  & 0.870 & 0.808 & 1.134 & 0.255 \\ 
  &  &  &  & (0.621) & (0.572) & (0.763) & (0.284) \\ 
  itusage\_internet &  &  &  & 0.688 & 0.422 & 0.727 & 0.044 \\ 
  &  &  &  & (0.515) & (0.468) & (0.600) & (0.229) \\ 
  itusage\_buy &  &  &  & 0.573 & 1.428$^{**}$ & 0.052 & 0.198 \\ 
  &  &  &  & (0.632) & (0.624) & (0.711) & (0.270) \\ 
  itusage\_excel &  &  &  & 0.612 & 0.717$^{*}$ & 1.080$^{*}$ & 0.224 \\ 
  &  &  &  & (0.471) & (0.430) & (0.565) & (0.226) \\ 
  itusage\_word &  &  &  & $-$0.886$^{*}$ & $-$0.714 & $-$1.117$^{*}$ & 0.028 \\ 
  &  &  &  & (0.517) & (0.479) & (0.641) & (0.228) \\ 
  itusage\_code &  &  &  & $-$1.780$^{**}$ & $-$1.282$^{*}$ & $-$2.234$^{**}$ & $-$0.251 \\ 
  &  &  &  & (0.795) & (0.771) & (0.910) & (0.423) \\ 
  itusage\_discuss &  &  &  & 2.559 & 2.909$^{*}$ & 3.612$^{*}$ & $-$0.163 \\ 
  &  &  &  & (1.604) & (1.654) & (1.939) & (0.505) \\ 
\end{tabular}}
\end{table}

\begin{table}[h!] \centering 
\scalebox{0.8}{
\begin{tabular}{@{\extracolsep{5pt}}lccccccc} 
\tiny
\\[-1.8ex]\hline 
\hline \\[-1.8ex] 
 & \multicolumn{7}{c}{\textit{Dependent variable}} \\ 
\cline{2-8} 
\\[-1.8ex] & Binary & Binary & Binary & Binary & Binary & Binary & Fractional \\ 
\\[-1.8ex] & (1) & (2) & (3) & (4) & (5) & (6) & (7)\\ 
\hline \\[-1.8ex] 
    physical\_long &  &  &  & $-$1.906$^{***}$ & $-$1.561$^{***}$ & $-$2.439$^{***}$ & $-$0.536$^{**}$ \\ 
  &  &  &  & (0.435) & (0.406) & (0.529) & (0.212) \\ 
  physical\_accurate &  &  &  & $-$0.014 & 0.138 & 0.183 & $-$0.059 \\ 
  &  &  &  & (0.387) & (0.356) & (0.471) & (0.174) \\ 
  planning\_own &  &  &  & $-$0.364 & $-$0.580 & $-$0.463 & $-$0.196 \\ 
  &  &  &  & (0.424) & (0.394) & (0.504) & (0.186) \\ 
  planning\_others &  &  &  & $-$0.431 & $-$0.459 & 0.193 & 0.014 \\ 
  &  &  &  & (0.576) & (0.520) & (0.692) & (0.271) \\ 
  planning\_time &  &  &  & $-$0.841$^{*}$ & $-$0.945$^{**}$ & $-$0.454 & $-$0.003 \\ 
  &  &  &  & (0.459) & (0.425) & (0.532) & (0.216) \\ 
  problem\_simple &  &  &  & 0.246 & 0.336 & 0.056 & $-$0.004 \\ 
  &  &  &  & (0.439) & (0.412) & (0.509) & (0.202) \\ 
  problem\_complex &  &  &  & $-$0.217 & $-$0.255 & 0.050 & $-$0.075 \\ 
  &  &  &  & (0.557) & (0.534) & (0.652) & (0.266) \\ 
  reading\_instruction &  &  &  & 0.865$^{*}$ & 0.946$^{**}$ & 0.801 & 0.085 \\ 
  &  &  &  & (0.465) & (0.422) & (0.520) & (0.201) \\ 
  reading\_letter &  &  &  & 0.792 & 0.399 & 0.165 & 0.234 \\ 
  &  &  &  & (0.723) & (0.666) & (0.849) & (0.334) \\ 
  reading\_news &  &  &  & $-$0.443 & $-$0.155 & 0.310 & $-$0.011 \\ 
  &  &  &  & (0.520) & (0.471) & (0.659) & (0.229) \\ 
  reading\_article &  &  &  & $-$0.665 & $-$0.720 & $-$0.969 & $-$0.047 \\ 
  &  &  &  & (0.781) & (0.718) & (0.940) & (0.322) \\ 
  reading\_book &  &  &  & $-$3.639$^{***}$ & $-$3.855$^{***}$ & $-$3.632$^{***}$ & $-$0.304 \\ 
  &  &  &  & (1.064) & (1.086) & (1.249) & (0.352) \\ 
  reading\_manual &  &  &  & $-$0.303 & $-$0.341 & $-$0.306 & 0.090 \\ 
  &  &  &  & (0.594) & (0.566) & (0.692) & (0.258) \\ 
  reading\_bill &  &  &  & 0.697 & 0.689 & 1.227$^{**}$ & 0.212 \\ 
  &  &  &  & (0.500) & (0.453) & (0.600) & (0.212) \\ 
  reading\_graph &  &  &  & 0.156 & 0.236 & 0.192 & $-$0.009 \\ 
  &  &  &  & (0.462) & (0.428) & (0.546) & (0.201) \\ 
  wricalc\_letter &  &  &  & $-$0.644 & $-$0.614 & $-$0.112 & $-$0.218 \\ 
  &  &  &  & (0.661) & (0.638) & (0.762) & (0.304) \\ 
  wricalc\_news &  &  &  & 2.055 & 3.158 & 4.665 & 0.148 \\ 
  &  &  &  & (3.602) & (2.784) & (4.245) & (1.017) \\ 
  wricalc\_report &  &  &  & $-$2.091$^{***}$ & $-$1.980$^{***}$ & $-$1.676$^{***}$ & $-$0.415$^{**}$ \\ 
  &  &  &  & (0.446) & (0.422) & (0.534) & (0.207) \\ 
  wricalc\_form &  &  &  & $-$0.169 & $-$0.135 & $-$0.225 & 0.138 \\ 
  &  &  &  & (0.401) & (0.380) & (0.464) & (0.185) \\ 
  wricalc\_budget &  &  &  & 0.006 & 0.016 & $-$0.473 & $-$0.034 \\ 
  &  &  &  & (0.511) & (0.467) & (0.604) & (0.228) \\ 
  wricalc\_fraction &  &  &  & 0.861$^{*}$ & 0.745 & 0.827 & 0.062 \\ 
  &  &  &  & (0.513) & (0.496) & (0.599) & (0.232) \\ 
  wricalc\_calculator &  &  &  & 2.111$^{***}$ & 2.097$^{***}$ & 1.893$^{***}$ & 0.478$^{**}$ \\ 
  &  &  &  & (0.441) & (0.413) & (0.531) & (0.224) \\ 
  wricalc\_chart &  &  &  & 0.363 & $-$0.060 & 0.740 & $-$0.079 \\ 
  &  &  &  & (0.695) & (0.648) & (0.809) & (0.287) \\ 
  wricalc\_simple &  &  &  & $-$0.408 & $-$0.578 & $-$0.535 & 0.032 \\ 
  &  &  &  & (0.533) & (0.517) & (0.640) & (0.237) \\ 
  wricalc\_advanced &  &  &  & 2.041 & 0.800 & 2.647 & $-$0.006 \\ 
  &  &  &  & (1.247) & (1.103) & (1.923) & (0.492) \\ 
  Constant & $-$3.708$^{*}$ & $-$7.828$^{***}$ & $-$11.294$^{***}$ & $-$6.391$^{**}$ & 0.154 & $-$6.814$^{*}$ & $-$0.768 \\ 
  & (1.895) & (2.088) & (2.726) & (2.783) & (0.551) & (4.092) & (1.805) \\ 
 \hline \\[-1.8ex] 
Observations & 541 & 541 & 541 & 541 & 541 & 541 & 995 \\ 
Akaike Inf. Crit. & 663.473 & 558.371 & 506.827 & 391.567 & 421.820 & 347.170 & 1,122.227 \\ 
\hline 
\hline \\[-1.8ex] 
\textit{Note:} & \multicolumn{7}{r}{$^{*}$p$<$0.1; $^{**}$p$<$0.05; $^{***}$p$<$0.01; \textit{Standard errors are shown in parentheses}} \\ 
\end{tabular}}
\end{table} 

\newpage
\begin{table}[h!] \centering 
\centering
\caption{When comparing the outcome reported in past contributions with our findings, the choice of model clearly dictates the resulting probabilities. Models with a binary dependent variable lead to bimodal distributions with large high-risk groups. Fractional models yield a normal distribution with small high-risk shares.}
\label{tab:comparison}
\scalebox{0.8}{ 
\begin{tabular}{l l l p{2cm} l c}
\\[-1.8ex]\hline 
\hline \\[-1.8ex] 
Author & Initial Input & Model Type & Predicted Distribution & High Risk & Country               \\
\hline 
\cite{frey2017future} & Binary (0/1) & Classification & Bimodal & 47\% & US \\
&&&&& \\
\cite{bowles2014computerisation} & \multicolumn{2}{c}{\textit{transfer of \cite{frey2017future}}} & Bimodal & 54\% & AT\\
\cite{arntz2016risk} & Discrete (0-1) & Fractional & Normal & 12\% & AT \\
\cite{nagl2017digitalisierung} & Discrete (0-1) & Fractional & Normal & 9\% & AT \\
&&&&& \\
\textit{Own calculations} & Binary (0/1) & Logit & Bimodal & 45\% & AT \\
\textit{Own calculations} & Discrete (0-1) & Fractional & Normal & 12\% & AT\\
\textit{Own calculations} & Binary (0/1) & LDA & Bimodal & 46\% & AT\\
\hline
\hline \\
\end{tabular}}
\end{table}

\newpage
\begin{landscape}
\begin{table}[!htbp] \centering 
  \caption{The correlation matrix shows that there are no significant correlations across variables, except in the case of the test scores for problem-solving (PS), numeracy and literacy.} 
  \label{tab:correlation} 
  \scalebox{0.7}{
\begin{tabular}{@{\extracolsep{5pt}} ccccccccccccc} 
\\[-1.8ex]\hline 
\hline \\[-1.8ex] 
 & Age Group & Gender & Education & Education$^{2}$ & Firm - Sector & Firm - Size & Skills - PS & Skills - Numeracy & Skills - Literacy & Job - Resp. & Job - Education & Job - Experience \\ 
\hline \\[-1.8ex] 
Age Group & $1$ & $$-$0.031$ & $0.120$ & $0.108$ & $$-$0.153$ & $0.014$ & $$-$0.298$ & $0.018$ & $$-$0.110$ & $0.167$ & $0.242$ & $0.244$ \\ 
Gender & $$-$0.031$ & $1$ & $0.001$ & $$-$0.001$ & $$-$0.134$ & $$-$0.169$ & $$-$0.122$ & $$-$0.183$ & $$-$0.051$ & $$-$0.220$ & $$-$0.013$ & $$-$0.196$ \\ 
Education & $0.120$ & $0.001$ & $1$ & $0.992$ & $$-$0.177$ & $0.101$ & $0.316$ & $0.418$ & $0.424$ & $0.138$ & $0.581$ & $0.197$ \\ 
Education$^{2}$ & $0.108$ & $$-$0.001$ & $0.992$ & $1$ & $$-$0.185$ & $0.104$ & $0.312$ & $0.414$ & $0.421$ & $0.135$ & $0.576$ & $0.194$ \\ 
Firm - Sector & $$-$0.153$ & $$-$0.134$ & $$-$0.177$ & $$-$0.185$ & $1$ & $$-$0.082$ & $0.088$ & $0.019$ & $$-$0.025$ & $0.021$ & $$-$0.198$ & $0.126$ \\ 
Firm - Size & $0.014$ & $$-$0.169$ & $0.101$ & $0.104$ & $$-$0.082$ & $1$ & $0.158$ & $0.116$ & $0.123$ & $0.071$ & $0.113$ & $0.108$ \\ 
Skills - PS & $$-$0.298$ & $$-$0.122$ & $0.316$ & $0.312$ & $0.088$ & $0.158$ & $1$ & $0.744$ & $0.810$ & $0.024$ & $0.209$ & $0.064$ \\ 
Skills - Numeracy & $0.018$ & $$-$0.183$ & $0.418$ & $0.414$ & $0.019$ & $0.116$ & $0.744$ & $1$ & $0.865$ & $0.096$ & $0.358$ & $0.141$ \\ 
Skills - Literacy & $$-$0.110$ & $$-$0.051$ & $0.424$ & $0.421$ & $$-$0.025$ & $0.123$ & $0.810$ & $0.865$ & $1$ & $0.058$ & $0.337$ & $0.105$ \\ 
Job - Resp. & $0.167$ & $$-$0.220$ & $0.138$ & $0.135$ & $0.021$ & $0.071$ & $0.024$ & $0.096$ & $0.058$ & $1$ & $0.161$ & $0.302$ \\ 
Job-Education & $0.242$ & $$-$0.013$ & $0.581$ & $0.576$ & $$-$0.198$ & $0.113$ & $0.209$ & $0.358$ & $0.337$ & $0.161$ & $1$ & $0.300$ \\ 
Job-Experience & $0.244$ & $$-$0.196$ & $0.197$ & $0.194$ & $0.126$ & $0.108$ & $0.064$ & $0.141$ & $0.105$ & $0.302$ & $0.300$ & $1$ \\ 
\hline \\[-1.8ex] 
\end{tabular} } 
\end{table} 
\end{landscape}

\end{document}